\documentclass[sigconf]{acmart}

\usepackage{booktabs} 
\usepackage{epsfig}
\usepackage{graphicx}
\usepackage{amsmath}
\usepackage{amssymb}
\usepackage{bm}

\usepackage{algorithm}
\usepackage{algorithmic}
\usepackage{url}
\usepackage{subfigure}
\usepackage{bbding}
\usepackage{multirow}
\usepackage{enumitem}
\usepackage{array}

\copyrightyear{2018}
\acmYear{2018}
\setcopyright{acmcopyright}
\acmConference[ICCSE'18]{The 3rd International Conference on Crowd Science and Engineering}{July 28--31, 2018}{Singapore, Singapore}
\acmPrice{15.00}
\acmDOI{10.1145/3265689.3265705}
\acmISBN{978-1-4503-6587-1/18/07}

\begin{document}
\title{Deep Transfer Learning for Cross-domain Activity Recognition}



%

%
\author{Jindong Wang}
\authornote{Jindong Wang and Yiqiang Chen are also with Beijing Key Lab of Mobile Computing and Pervasive Devices and University of Chinese Academy of Sciences.}
\affiliation{%
  \institution{Institute of Computing Technology, CAS}
  \country{Beijing, China}}
\email{wangjindong@ict.ac.cn}
\author{Vincent W. Zheng}
\affiliation{
  \institution{Advanced Digital Sciences Center
  	}
  \country{Singapore}}
\email{vincent.zheng@adsc.com.sg}
\author{Yiqiang Chen}
\authornote{Corresponding author.}
\affiliation{%
	\institution{Institute of Computing Technology, CAS}
	\country{Beijing, China}}
\email{yqchen@ict.ac.cn}

\author{Meiyu Huang}
\affiliation{%
	\institution{Qian Xuesen Laboratory of Space Technology, CAST}
	\country{Beijing, China}}
\email{huangmeiyu@qxslab.cn}
%
%
%


\begin{abstract}
Human activity recognition plays an important role in people's daily life. However, it is often expensive and time-consuming to acquire sufficient labeled activity data. To solve this problem, transfer learning leverages the labeled samples from the source domain to annotate the target domain which has few or none labels. Unfortunately, when there are several source domains available, it is difficult to select the \textit{right} source domains for transfer. The right source domain means that it has the most similar properties with the target domain, thus their similarity is higher, which can facilitate transfer learning. Choosing the right source domain helps the algorithm perform well and prevents the negative transfer. In this paper, we propose an effective \textit{Unsupervised Source Selection algorithm for Activity Recognition (USSAR)}. USSAR is able to select the most similar $K$ source domains from a list of available domains. After this, we propose an effective \textit{Transfer Neural Network} to perform knowledge transfer for Activity Recognition (TNNAR). TNNAR could capture both the time and spatial relationship between activities while transferring knowledge. Experiments on three public activity recognition datasets demonstrate that: 1) The USSAR algorithm is effective in selecting the best source domains. 2) The TNNAR method can reach high accuracy when performing activity knowledge transfer.
\end{abstract}

%
%

\ccsdesc[500]{Human-centered computing~Ubiquitous computing}
\ccsdesc[500]{Computing methodologies~Transfer learning}

\keywords{Transfer Learning, Activity Recognition, Deep Learning, Domain Adaptation}

\maketitle

\section{Introduction}

Human activity recognition~(HAR) aims to seek the profound high-level information from the low-level sensor inputs~\cite{wang2017deep}. For example, we can predict if a person is walking or running using the on-body sensors such as a smartphone or wristband. HAR has been widely applied to many applications such as indoor localization~\cite{xu2016indoor}, sleep state detection~\cite{zhao2017sleep}, and smart home sensing~\cite{wen2016adaptive}. 

\begin{figure}[t!]
	\centering
	\includegraphics[scale=0.45]{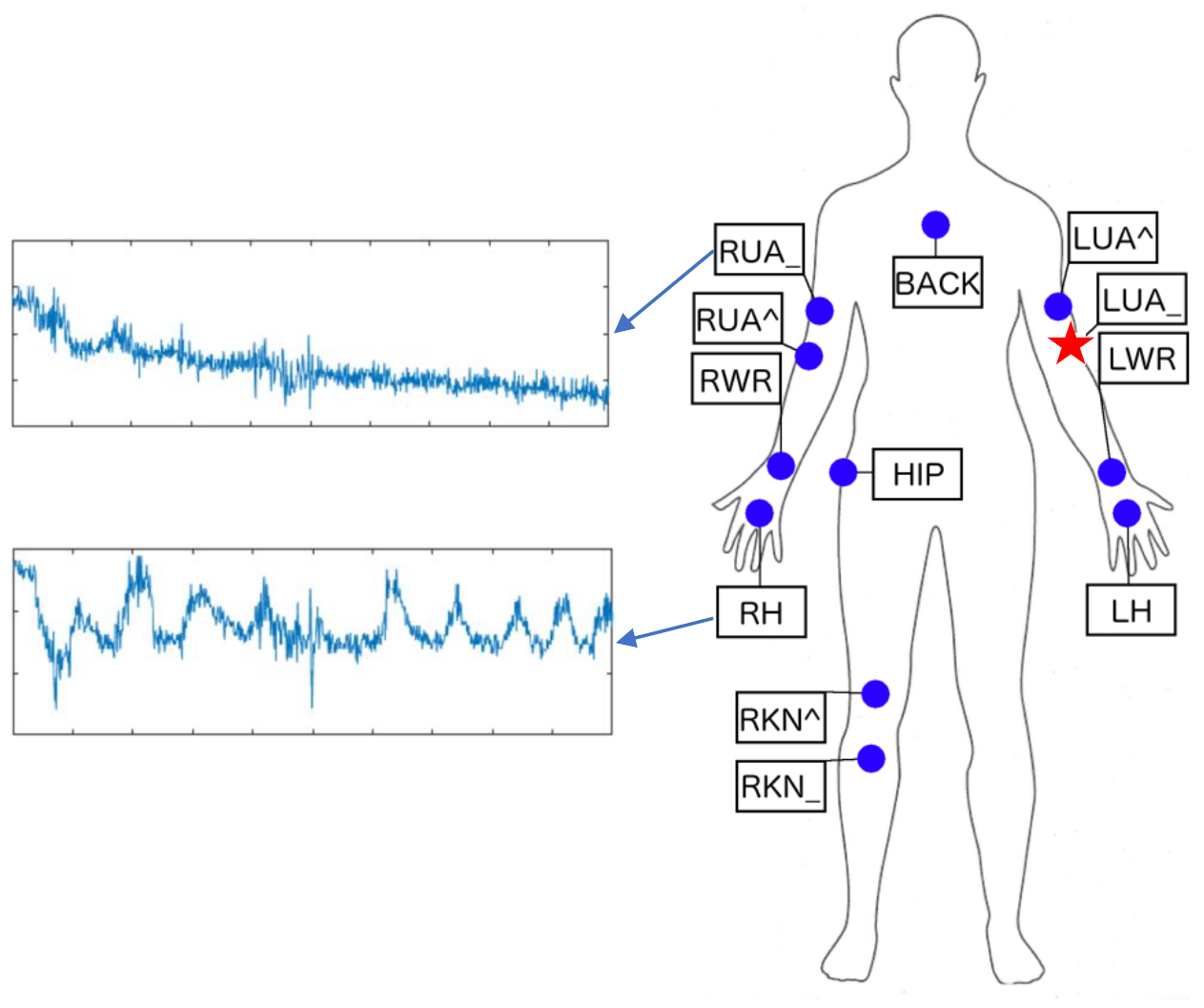}
	\vspace{-.1in}
	\caption{An example of cross-domain activity recognition. The activity signals on different body parts are often different. If the labels of a certain part are missing (the red pentacle), how to leverage the well-labeled activity data on other body parts (the blue dots) to acquire its labels?}
	\label{fig-intro}
	\vspace{-.2in}
\end{figure}

There are different activity patterns of different body parts, so sensors can be put on them to collect activity data and then build machine learning models. The combination of signals from different body parts can be used to reflect meaningful knowledge such as person's detailed health information~\cite{hammerla2015pd} and working states~\cite{plotz2011feature}. Unfortunately, the real situation is that we either do not want all body parts to be equipped with sensors, or the data on certain body part may be easily missing. In these situations, we are unable to train models to recognize the activities of some body parts. Figure~\ref{fig-intro} illustrates this situation. If the activity information on the arm (the red pentacle, we call it the \textit{target domain}) is missing and we want to predict the activities on this body part, how to utilize the information on other parts (such as torso or leg, we call them the \textit{source domains}) to help build the model? This problem is referred to as the \textit{cross-domain activity recognition (CDAR).}

\begin{figure}[t!]
	\centering
	\includegraphics[scale=0.45]{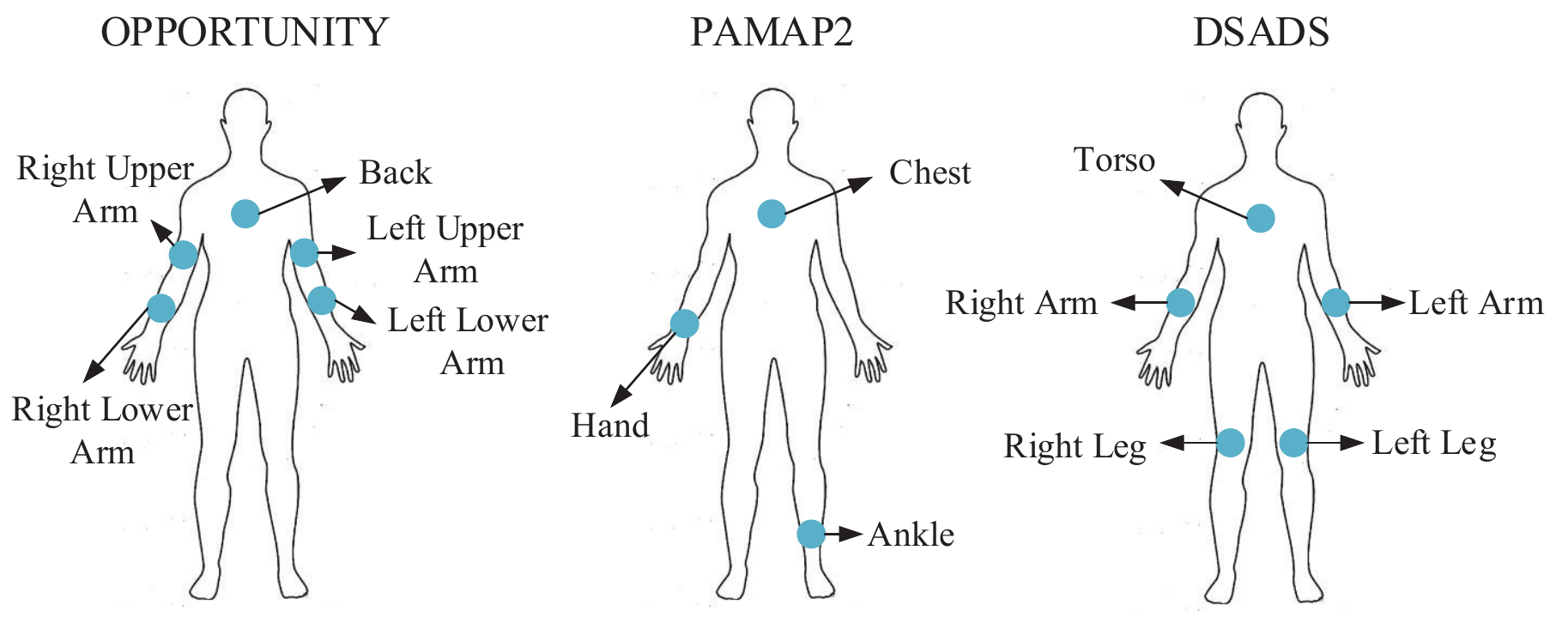}
	\vspace{-.1in}
	\caption{How to select the most similar body parts and perform activity transfer on multiple persons?}
	\label{fig-dataset}
	\vspace{-.2in}
\end{figure}

This problem is extremely challenging. Firstly, we do not know which body parts are most similar to the target domain since the sensor signals are not independent, but are highly correlated because of the shared body structures and functions. If we use all the body parts, there is likely to be negative transfer~\cite{pan2010survey} because some body parts may be dissimilar. Secondly, we only have the raw activity data on the target domain without the actual activity labels, making it harder to measure the similarities. Thirdly, even we already know the similar body parts to the target domain, it is also difficult to build a good machine learning model using both the source and target domains. The reason is that the signal on different domains are following exactly different distributions, which means there are distribution discrepancies between them. However, traditional machine learning models are built by assuming that all signals are with the same distribution. Moreover, existing neural networks are too generic to capture both the time and spatial information in the activity data. Fourthly, when it comes to different persons, there are also similar body parts across persons~(Fig.~\ref{fig-dataset}). This makes the problem more challenging.

In this paper, we aim to tackle above challenges through multiple source selection and deep neural network. For multiple source selection, we calculate the distance between available sources and the target domains to select the most similar source domains. Our intuition is that the sensor signals may consist of generic and specific relationships about the body parts: the generic relationship means the data distance between two signals such as the Euclidean distance or cosine similarity; the specific relationship refers to the similar moving patterns or body functions between two body parts. By calculating these two significant distances, we can correctly measure the distances between different body parts, and thus select the \textit{right} source domains. Our algorithm does not depend on the availability of the target domain labels. We call this algorithm \textit{Unsupervised Source Selection for Activity Recognition (USSAR)}.

After obtaining the \textit{right} source domains via the USSAR algorithm, we propose a \textit{Transfer Neural Network for Activity Recognition (TNNAR)}. TNNAR is an end-to-end neural network to perform knowledge transfer across different domains. The important thing is that in order to reduce the distance between two domains, we add an \textit{adaptation} layer in the network to calculate the adaptation loss, which can be optimized jointly with the classification loss. The main structure of TNNAR is two convolutional blocks with max-pooling operations, one LSTM layer, and two fully-connected layers. We use the convolutional layers to extract the spatial features from the original activity data. The LSTM layer is mainly for capturing the time features~\cite{wang2017deep}. The fully-connected layers are used for nonlinear transformation. Finally, a softmax function is applied for classification. 

To validate the performance of our USSAR and TNNAR algorithms, we conduct extensive experiments on three large-scale public activity recognition datasets: OPPORTUNITY, PAMAP2, and UCI DSADS. All of them contains multiple domains from several body parts. We perform cross-domain activity recognition across different body parts. The Experimental results demonstrate that USSAR is effective in choosing the \textit{right} source domains, and TNNAR achieves the best accuracy in activity transfer.

The main contributions of this paper are summarized as follows:

1) We propose the \textit{first} unsupervised source selection algorithm for activity recognition. USSAR measures both the general and specific characteristics of activity information, hence it is capable of capturing the profound relationship between different domains.

2) We propose an end-to-end transfer neural network for cross-domain activity recognition. Different from existing deep transfer learning methods that need to extract features from human knowledge, our TNNAR can simultaneously perform classification and adaptation between two activity domains on the original data.

3) We evaluate the proposed USSAR and TNNAR algorithms in extensive experiments on cross-position activity recognition. Therefore, our algorithms can be applied to many applications to increase the generalization ability of activity models.

\section{Related Work}
\label{sec-related}

\subsection{Activity Recognition}
Human Activity recognition has been a popular research topic in pervasive computing~\cite{bulling2014tutorial} for its competence in learning profound high-level knowledge about human activity from raw sensor inputs. Several survey articles have elaborated the recent advance of activity recognition using conventional machine learning~\cite{bulling2014tutorial,lara2013survey} and deep learning~\cite{wang2017deep} approaches.

Conventional machine learning approaches have made tremendous progress on HAR by adopting machine learning algorithms such as similarity-based approach~\cite{zheng2011user,chen2016ocean}, active learning~\cite{hossain2016active}, crowdsourcing~\cite{lasecki2013real}, and other semi-supervised methods~\cite{nguyen2015did,hu2016less}. Those methods typically treat HAR as a standard time series classification problem. And they tend to solve it by subsequently performing preprocessing procedures, feature extraction, model building, and activity inference. However, they all assume that the training and test data are with the same distribution. As for CDAR where the training~(source) and the test~(target) data are from different feature distributions, those conventional methods are prune to under-fitting since their generalization ability will be undermined~\cite{pan2010survey}.

Deep learning based HAR~\cite{wang2017deep,plotz2011feature} achieves the state-of-the-art performance than conventional machine learning approaches. The reason is that deep learning is capable of automatically extracting high-level features from the raw sensor readings~\cite{lecun2015deep}. Therefore, the features are likely to be more domain-invariant and tend to perform better for cross-domain tasks. A recent work evaluated deep models for cross-domain HAR~\cite{morales2016deep}, which provides some experience for future research on this area. There are still many open problems for deep learning based CDAR.

\subsection{Transfer Learning}
Transfer learning has been successfully applied in many applications such as Wi-Fi localization~\cite{pan2008transfer}, natural language processing~\cite{blitzer2006domain}, and visual object recognition~\cite{duan2012domain}. According to the literature survey~\cite{pan2010survey}, transfer learning can be categorized into 3 types: instance-based, parameter-based, and feature-based methods. 

Instance-based methods perform knowledge transfer mainly through instance re-weighting techniques~\cite{tan2017distant,chattopadhyay2012multisource}. Parameter-based methods~\cite{yao2010boosting,zhao2011cross} first train a model using the labeled source domain, then perform clustering on the target domain. Feature-based methods~\cite{pan2011domain,gong2012geodesic,wang2017balanced} learn a feature transformation between domains when the distance can be minimized. A detailed survey on transfer learning for activity recognition is conducted in~\cite{cook2013transfer}.

\textit{Our work differs from transfer learning in the following two aspects:}

\textbf{1) Source selection}

The work~\cite{xiang2011source} first proposed a source-selection free transfer learning approach. They choose the source samples that are close to the target samples using the Laplacian Eigenmap. The work \cite{lu2014source} followed this idea in the text classification. However, both of them only focused on the sample selection, while our USSAR focuses on the selection of the whole domain. Collier et al.~\cite{collier2018cactusnets} investigated the transfer performance of different layers of a neural network in a grid search manner. But they did not perform source selection. The work~\cite{sung2017learning} developed a relation network, which can be used to evaluate the distance between different image samples. Yet they still focused on the single sample. Authors in \cite{bhatt2016multi} proposed a greedy multi-source selection algorithm. This selection algorithm could iteratively select the best $K$ source domains and then perform transfer learning based on this selection. However, their method is too general to focus on the domain-specific features of activity recognition.

\textbf{2) Transfer network}

In recent years, the deep transfer learning methods have dramatically increased the learning performance of transfer learning tasks. Some popular deep transfer learning methods include Deep Domain Confusion (DDC)~\cite{tzeng2014deep}, Joint Adaptation Network~(JAN)~\cite{long2017deep}, and Domain adversarial Neural Network~(DANN)~\cite{ganin2016domain}. They are typically based on deep neural networks, where they add an adaptation layer to reduce the distribution divergence between domains. Compared to existing deep transfer learning methods, our TNNAR network is tailored according to the characteristics of activity recognition. Hence, we not only consider the generic transfer learning scenario, but also focus on the spatial and time relationship between activities.

\section{Our Method}

In this section, we describe our Unsupervised Source Selection and Transfer Neural Network for activity recognition. Before that, we will first formally give the problem definition.

\subsection{Problem definition}
Assume we have an activity domain $\mathcal{D}_t=\{\mathbf{x}^j_t\}^{n_t}_{j=1}$ as the target domain which we want to learn its corresponding activity labels $\mathbf{y}_t$, i.e. to predict the person's activity state based on the sensor signals. Suppose we have $C$ activity states (labels). There are $M$ labeled source domains available: $\{\mathcal{D}^i_s\}^{M}_{i=1}$. Each source domain $\mathcal{D}^i_s=\{\mathbf{x}^j_s,y^j_s\}^{n^i_s}_{j=1}$. Note that the data distributions are not the same, i.e. $P(\mathbf{x}_s) \ne P(\mathbf{x}_t)$. We need to design algorithms to: 1) select the best $K (K < M)$ source domains (we denote them as $\mathcal{D}_s(K)$), and 2) perform effective transfer learning from $\mathcal{D}_s(K)$ to $\mathcal{D}_t$ in order to obtain $\mathbf{y}_t$.

\subsection{Unsupervised Source Selection}
\label{sec-source}
Since the target domain has no labels, it is challenging to measure the distance between $\mathcal{D}_t$ and $\mathcal{D}^i_s$. Moreover, the activity signals are not only normal time series data, there are more mixed information and relationships in different domains, such as the correlated body functions and patterns. Existing distance measurements such as Maximum Mean Discrepancy (MMD)~\cite{borgwardt2006integrating} and $\mathcal{A}$ distance~\cite{ben2007analysis} are too generic for our problem. They only calculate the data distance without considering the relationship between body parts. Therefore, we should develop a comprehensive measurement to evaluate the distance between different activity domains.

In this paper, we propose the Unsupervised Source Selection for Activity Recognition (USSAR) algorithm to effectively select the \textit{right} source domains for the target activity domain. Our USSAR well considers the \textit{generality} and \textit{specificity} of activity while selecting source domains. Generality means that we have to seek the general relation between activity data, which is a common problem in machine learning. More importantly, specificity means that we should consider the specific information behind different activity domains. Formally, if we denote $D(A,B)$ as the distance between domains $A$ and $B$, then it can be represented as

\begin{equation}
	\label{eq-general}
	D(A,B) = D_g(A,B) + \lambda D_s(A,B),
\end{equation}
where $D_g(A,B)$ is the general distance ($g$ for general) and $D_s(A,B)$ is the specific distance ($s$ for specific). $\lambda \in [0,1]$ is the trade-off factor between two terms.

The \textit{general} distance $D_g(A,B)$ can be easily computed by the well-established $\mathcal{A}$-distance~\cite{ben2007analysis}:

\begin{equation}
	\label{eq-dist-general}
	D_g(A,B) = 2(1-2\epsilon),
\end{equation}
where $\epsilon$ is the error to classify domains $A$ and $B$. In order to obtain $\epsilon$, we train a linear binary classifier $h$ on $A$ and $B$, where $A$ has the label $+1$ and $B$ has the label $-1$ (or vice versa). Then, we apply prediction on both domains to get the error $\epsilon$.

The \textit{specific} distance $D_s(A,B)$ is composed of two important aspects from activity recognition: the \textbf{semantic} and the \textbf{kinetic} information. The semantic information refers to the spatial relationship between two domains, i.e. close spatial relations lead to the similar property. The kinetic information refers to the activity signal relationship between domains, i.e. close signal relations lead to the shorter distance. We take the following approaches to calculate the semantic and kinetic distances. 

\textbf{Semantic} distance: We basically give each source domain a weight $w \in [0,1]$ indicating its spatial relationship with the target domain. Therefore, if there are $M$ source domains available, there will be $M$ weights: $\{w_i\}^M_{i=1}$. Currently, the weighting technique is only based on human experience, i.e. we give relatively small weights if we think two domains are not closely related, while we give relatively large weights if we think two domains are closely related. For instance, if the target domain is the \textit{Right Hand}, we would probably give a larger weight to the source domain \textit{Left Hand} since these two body parts are always correlated; on the other hand, we would probably give the source domain \textit{Torso} a small weight since these two body parts are not exactly correlated. Since $w_i$ is a weight, we bound it by:

\begin{equation}
	\sum_{i=1}^{M} w_i = 1.
\end{equation}

\textbf{Kinetic} distance: There are many approaches available to approximate the signal relationship between two domains. This relationship can be captured by the consistency between two signals. For instance, the Pearson correlation score and the maximum mean discrepancy can be used to calculate this relationship. In this paper, we adopt the well-established \textit{Cosine} similarity function. Specifically, given two domains $A$ and $B$, their cosine similarity is formulated as:

\begin{equation}
	cos(A,B) = \mathbb{E}\left[\sum_{\mathbf{a},\mathbf{b}} \frac{\mathbf{a} \cdot \mathbf{b}}{|\mathbf{a} \cdot \mathbf{b}|}\right],
\end{equation}
where $\mathbf{a},\mathbf{b}$ are the basic vectors in $A$ and $B$, respectively. $\mathbb{E}[\cdot]$ is the expectation of samples. Note that when one domain is the target domain, its weight can be set to 1. Once the kinetic distance is calculated, we combine it with the weights generated by the semantic distance, and finally get the specificity distance as:

\begin{equation}
	\label{eq-dist-specific}
	d_s(A,B) = \mathbb{E}\left[\sum_{\mathbf{a},\mathbf{b}} \frac{w_{\mathbf{a}}\mathbf{a} \cdot w_{\mathbf{b}}\mathbf{b}}{|w_{\mathbf{a}}\mathbf{a} \cdot w_{\mathbf{b}}\mathbf{b}|}\right].
\end{equation}

\textit{Overall distance}: we combine the general distance~(Eq. \ref{eq-dist-general}) and the specific distance~(Eq. \ref{eq-dist-specific}) to get the final distance expression:

\begin{equation}
\label{eq-dist-ussar}
	d(A,B) = 2(1-2\epsilon) + \lambda \mathbb{E}\left[\sum_{\mathbf{a},\mathbf{b}} \frac{w_{\mathbf{a}}\mathbf{a} \cdot w_{\mathbf{b}}\mathbf{b}}{|w_{\mathbf{a}}\mathbf{a} \cdot w_{\mathbf{b}}\mathbf{b}|}\right].
\end{equation}

Note that this distance measurement does not rely on the labels of the target domain. We call this distance the \textit{Context Activity Distance (CAD)}. Once we have the CAD, we can perform source selection from many available source domains. In this paper, we propose a greedy algorithm to select the available sources. We regard the source selection problem as the process of constructing a finite set $S$. Initially, we calculate all the CADs between the target domain and every source domain and sort them in an increasing order. At this time, the set $S = \phi$. Then, we add the source domain with the smallest CAD (denoted as $\mathcal{D}^{\min}_s$)to $S$: $S={\mathcal{D}^{\min}_s}$. After that, we add other $K-1$ source domains to $S$ according to the greedy technique: if $d(S,\mathcal{D}^i_s) < d(\mathcal{D}_t,\mathcal{D}^i_s)$, we add $\mathcal{D}^i_s$ to $S$. The USSAR algorithm is illustrated in Algorithm~\ref{algo-ussar}.

\begin{figure*}[ht]
	\centering
	\includegraphics[scale=0.52]{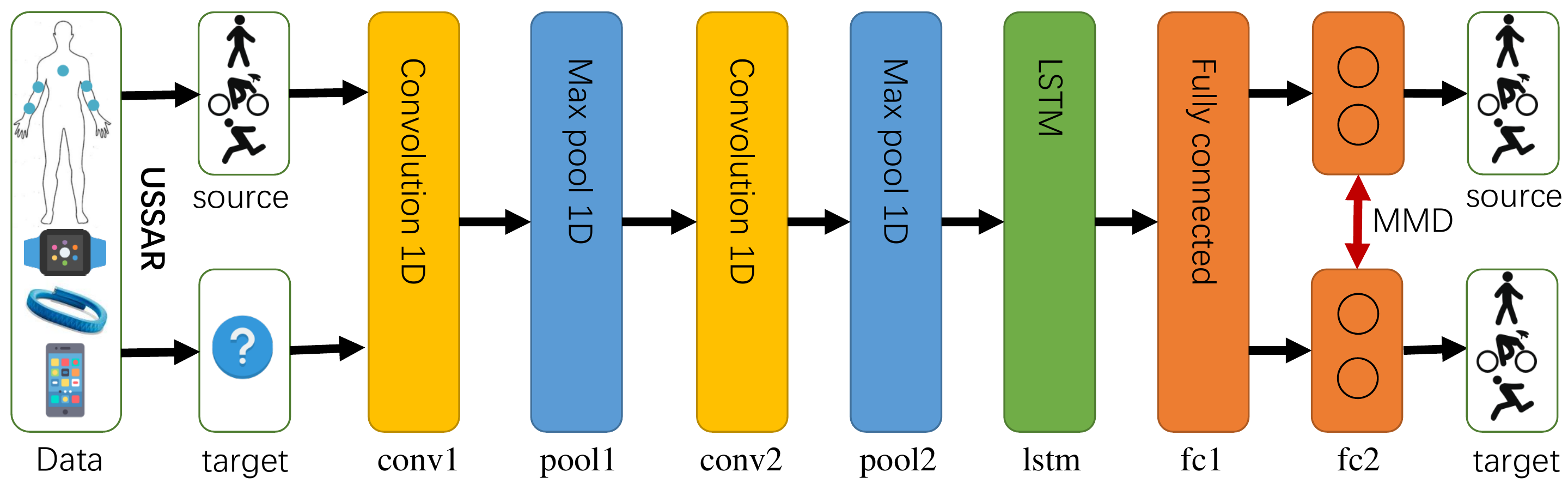}
	\vspace{-.1in}
	\caption{The structure of Transfer Neural Network for Activity Recognition (TNNAR).}
	\label{fig-net}
	\vspace{-.1in}
\end{figure*}

\begin{algorithm}[t!]
	\caption{USSAR: Unsupervised Source Selection for Activity Recognition}
	\label{algo-ussar}
	\renewcommand{\algorithmicrequire}{\textbf{Input:}} 
	\renewcommand{\algorithmicensure}{\textbf{Output:}} 
	\begin{algorithmic}[1]
		\REQUIRE~~
		$M$ available source domains $\{\mathcal{D}^i_s\}^M_{i=1}$, target domain $\mathcal{D}_t$, the number of selected source domains $K(K << M)$\\
		\ENSURE~~
		The selected source domain set $S$.\\
		\STATE Calculate the CAD between each source domain and target domain using Eq. (\ref{eq-dist-ussar}), and sort them in an increasing order;
		\STATE Initialize a set $S = \mathcal{D}^{\min}_s$;\\
		Set $i = 2$\\
		\REPEAT
		\STATE Calculate the CAD between $\mathcal{D}^i_s$ and $S$, and denote it as $d_{iS}$;
		\STATE If $d_{iS} < d(\mathcal{D}_t,\mathcal{D}^i_s)$, we add $\mathcal{D}^i_s$ to $S$;
		\STATE Else $i = i + 1$;
		\UNTIL{i = K}
		\RETURN $S$.
	\end{algorithmic}
\end{algorithm}

\subsection{Transfer Neural Network}
\label{sec-transfer}

After obtaining the selected source domains $\mathcal{D}_s(K)$, we can perform knowledge transfer across the target and the source domains. In this paper, we propose a neural network to accomplish this transfer. Generally speaking, the goal of the transfer neural network is to learn the classifiers $y = f_s(\mathbf{x})$ and $y = f_t(\mathbf{x})$ and minimize their discrepancies. So the expected target risk is bounded: $R_t(f_t)=\mathbb{E}_{(\mathbf{x},y)}[f_t(\mathbf{x}) \ne y]$. Since there are discrepancies between the source and target domain, the general form of the function $f$ can be expressed as:

\begin{equation}
	f(\mathbf{x}) = \ell_c(\mathbf{x}) + \mu \ell_a(\mathcal{D}_s,\mathcal{D}_t),
\end{equation}
where $\ell_c(\mathbf{x})$ is the classification loss on the labeled data (source domain), and $\ell_a(\cdot,\cdot)$ is the adaptation loss on both of the source and target domains. $\mu \in [0,1]$ is the trade-off factor.

This is a general form of a transfer neural network. In cross-domain activity recognition setting, the structure of the neural network has to be modified according to the characteristics of activity recognition. In this paper, we propose Transfer Neural Network for Activity Recognition (TNNAR). The structure of our TNNAR is illustrated in Figure~\ref{fig-net}. The proposed TNNAR consists of two convolutional layers with max-pooling layers, one LSTM layer, and two fully-connected (fc) layers. The convolutional layers are adopted to extract the spatial features for the activities, while the LSTM layer is adopted to capture the time relationship between activities. The fully-connected layer is acting as the classification function.

Beyond this simple structure, we also add an \textit{adaptation} layer to reduce the discrepancy between two domains. The reason we add this layer after the first fully-connected layer is that the features are becoming specific to the higher layer, making adaptation more urgent. The low-level layers are only extracting some common features, thus they do not need to be adapted~\cite{yosinski2014transferable}. The adaptation layer computes the adaptation loss, which can be optimized jointly with the classification loss.

We denote $\mathbf{W}^l$ and $\mathbf{b}^l$ the weights and bias at fc layer $l$. Each fc layer $l$ will learn a nonlinear mapping $\mathbf{h}^l = f^l(\mathbf{W}^l \mathbf{h}^{l-1}_i + \mathbf{b}^l)$, where $\mathbf{h}^l$ is the $l$th layer hidden representation and $f^l$ is the activation function. The activation $f^l$ for the last fc layer is computed as $f^l(\mathbf{x})=e^\mathbf{x} / \sum_{j=1}^{|\mathbf{x}|}e^{\mathbf{x}_j}$. The other fc layer takes the ReLU units $f^l=\max(\mathbf{0},\mathbf{x})$. If we denote $\bm{\Theta}=\{\mathbf{W}^l,\mathbf{b}^l\}$ the hyperparameters of the neural network, then the empirical risk is

\begin{equation}
\ell_c = \min_{\bm{\Theta}} \frac{1}{n_b} \sum_{i=1}^{b} J(\bm{\Theta}(\mathbf{x}^b_i),y^b_i),
\end{equation}
where $J$ is the cross-entropy function. $(\mathbf{x}^b_i,y^b_i)$ denotes all the labeled samples from the source domain.

As for the adaptation layer, we adopted the well-established Maximum Mean Discrepancy (MMD)~\cite{borgwardt2006integrating} as the measurement to reduce the discrepancy between domains. MMD is a popular distance metric, which has been widely used in many existing work~\cite{pan2011domain,long2013transfer}, and its effectiveness has been verified in~\cite{gretton2012kernel}. The MMD distance between distributions $p$ and $q$ is defined as $d^2(p,q)=(\mathbb{E}_p[\phi(\mathbf{z}_s)] - \mathbb{E}_q[\phi(\mathbf{z}_t)])^2_{\mathcal{H}_{K}}$ where $\mathcal{H}_{K}$ is the reproducing kernel Hilbert space (RKHS) induced by feature map $\phi(\cdot)$. Here, $\mathbb{E}[\cdot]$ denotes the mean of the embedded samples. Therefore, the MMD distance between the source and target domain is

\begin{equation}
\label{eq-mmd}
	MMD(\mathcal{D}_s,\mathcal{D}_t) = \Vert\mathbb{E}[\mathbf{x}_s] - \mathbb{E}[\mathbf{x}_t]\Vert^2_{\mathcal{H}_{K}}.
\end{equation}

We train the TNNAR using mini-batch Stochastic Gradient Descent (SGD) strategy. The gradient can be calculated as

\begin{equation}
	\Delta_{\bm{\Theta}^l} = \frac{\partial J(\cdot)}{\partial \bm{\Theta}^l} + \mu \frac{\partial \ell_a(\cdot)}{\partial \bm{\Theta}^l}.
\end{equation}

\section{Experimental Evaluation}
\label{sec-exp}

In this section, we evaluate the performance of USSAR and TNNAR.

\subsection{Datasets and Setup}

We used the same datasets and setup in a recent literature~\cite{wang2018stratified} to perform cross-position activity recognition (CPAR). CPAR is an important aspect of cross-domain activity recognition (CDAR). Specifically, it refers to the situation where the activity labels of some body parts are missing, so it is necessary and feasible to leverage the labeled data from other similar body parts to get the labels of those body parts. 

\begin{table*}[htbp]
	\centering
	\caption{Brief introduction of three public datasets for activity recognition~\cite{wang2018stratified}}
	\vspace{-.1in}
	\label{tb-dataset}
	\resizebox{\textwidth}{!}{
		\begin{tabular}{|c|c|c|c|c|c|}
			\hline
			\textbf{Dataset} & \textbf{\#Subject} & \textbf{\#Activity} & \textbf{\#Sample} & \textbf{\#Feature} & \textbf{Body parts} \\ \hline \hline
			OPPORTUNITY & 4 & 4 & 701,366 & 459 & Back (B), Right Upper Arm (RUA), Right Left Arm (RLA), Left Upper Arm (LLA), Left Lower Arm (LLA) \\ \hline
			PAMAP2 & 9 & 18 & 2,844,868 & 243 & Hand (H), Chest(C), Ankle (A) \\ \hline
			DSADS & 8 & 19 & 1,140,000 & 405 & Torso (T), Right Arm (RA), Left Arm (LA), Right Leg (RL), Left Leg (LL) \\ \hline
		\end{tabular}
	}
	\vspace{-.2in}
\end{table*}

There are three public datasets used in~\cite{wang2018stratified}: OPPORTUNITY dataset~(\textbf{OPP})~\cite{chavarriaga2013opportunity}, PAMAP2 dataset~(\textbf{PAMAP2})~\cite{reiss2012introducing}, and UCI daily and sports dataset~(\textbf{DSADS})~\cite{barshan2014recognizing}. Table~\ref{tb-dataset} provides a brief introduction to these three datasets. In the following, we briefly introduce those datasets, and more information can be found in their original papers. \textbf{OPP} is composed of 4 subjects executing different levels of activities with sensors tied to more than 5 body parts. \textbf{PAMAP2} is collected by 9 subjects performing 18 activities with sensors on 3 body parts. \textbf{DSADS} consists of 19 activities collected from 8 subjects wearing body-worn sensors on 5 body parts. Accelerometer, gyroscope, and magnetometer are all used in three datasets.

Following the same protocol in~\cite{wang2018stratified}, we also investigated the same positions in three datasets as in Figure~\ref{fig-dataset}. In our experiments, we use the data from all three sensors in each body part since most information can be retained in this way.

The transfer scenarios are obtained according to~\cite{wang2018stratified}. There are three scenarios that reflect different similarities between domains: a)~\textit{similar body parts of the same person}, b)~\textit{different body parts of the same person}, c)~\textit{similar body parts of different person}. In the sequel, we use the notation $A \rightarrow B$ to denote labeling the activity of domain $B$ using the labeled domain $A$. In total, we constructed 22 tasks. Note that there are different activities in three datasets. For scenario a) and b), we simply use all the classes in each dataset; for scenario c)~which is cross-dataset, we extract 4 common classes for each dataset~(i.e. \textit{Walking, Sitting, Lying}, and \textit{Standing}). In addition, we did not include the scenario `\textit{different body parts of different person}' since 1)~all the methods perform poorly in that scenario, and 2)~that scenario does not have reasonable feasibility in real applications.

\begin{figure*}[htbp]
	\centering
	\subfigure[Torso as the target domain]{
		\centering
		\includegraphics[scale=0.45]{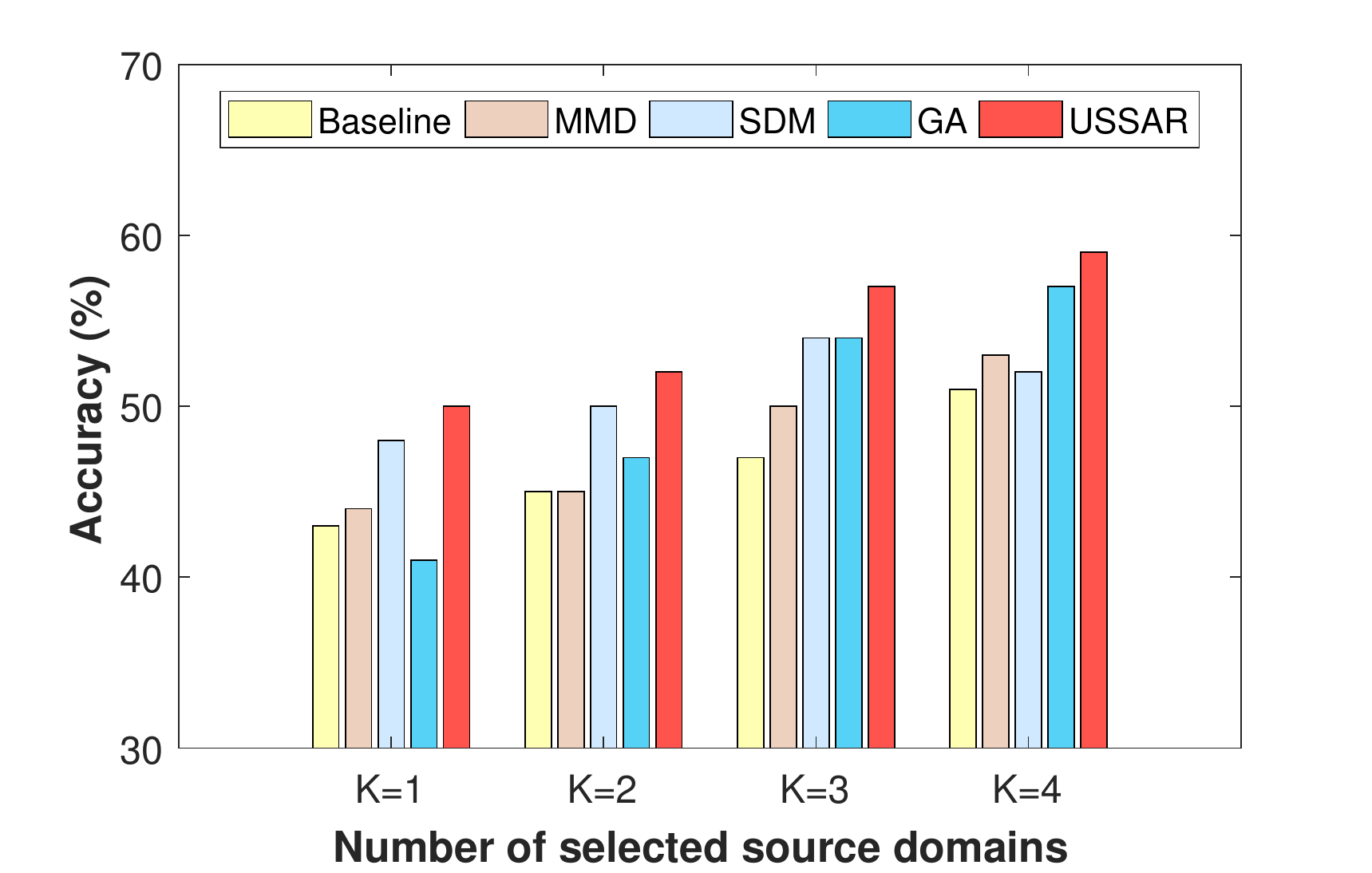}
		\label{fig-sub-sourcea}}
	\subfigure[Right Arm as the target domain]{
		\centering
		\includegraphics[scale=0.45]{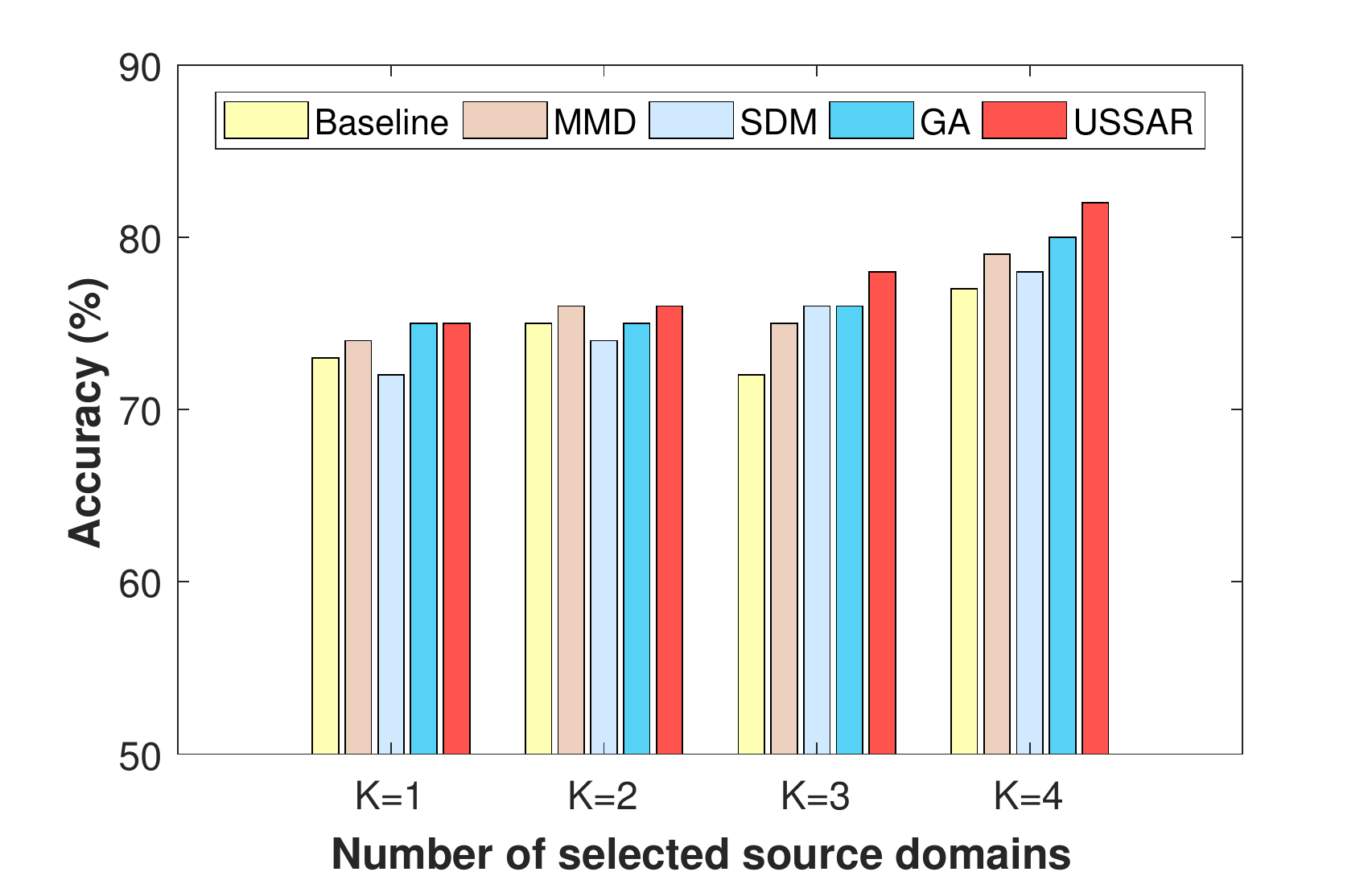}
		\label{fig-sub-sourceb}}
	\vspace{-.2in}
	\caption{Classification accuracy of different source selection algorithms.}
	\label{fig-select}
	\vspace{-.2in}
\end{figure*}

\subsection{Evaluation of USSAR}

First, we evaluate the performance of the proposed USSAR algorithm for unsupervised source selection. In this experiment, we choose the OPP and DSADS datasets since there are 5 body positions in them. We combine the two datasets together, which means that given a target domain in a dataset, there are 9 source domains available to select. In order to fully evaluate the algorithm, we set $K={1,2,3,4}$. 

We compare USSAR with several source selection techniques:

\begin{itemize}
	\item $\mathcal{A}$-distance~\cite{ben2007analysis}, which is selected according to the top $K$ smallest $\mathcal{A}$-distances. This distance is acting as the baseline.
	\item SDM: subspace disagreement measurement~\cite{gong2012geodesic}, which is computing the distance between domains based on the principle angle~\cite{hamm2008grassmann}.
	\item GR: Greedy algorithm~\cite{bhatt2016multi}, which is selecting sources using a greedy technique.
	\item MMD: Maximum Mean Discrepancy~\cite{borgwardt2006integrating}, which is a popular metric to measure the distance.
\end{itemize}

We randomly select two body positions: \textit{Torso (T, or Back in DSADS)} and \textit{Right Arm (RA) in DSADS}. When one of the body parts is selected as the target domain, it means that their labels are missing and we need to predict its labels using the rest 9 domains. The parameters of comparison methods are set according to their original papers. For USSAR algorithm, we assign different weights to each body part according to their relationship to the target domain. In fact, each body part can be a target domain, so these weights are dynamic and relative. For instance, in OPP dataset, if we take the RUA as the target domain, then the weights for the other nine parts can be: $0.1,0.2,0.3,0.1,\cdots,0.1$, as long as they add up to 1. This weighting technique can be tailored according to human experience. As long as it reflects the functional relationship between body parts, that can be accepted.

Note that there is no ground-truth about the \textit{right} source domains: we can never learn the actual distance between two domains. Therefore, we turn to use the classification accuracy as the evaluation metric. After selecting the source domains, we train the same linear SVM classifier on the selected source domains, and then apply prediction on the target domain to get the classification accuracy. It is intuitive that if we select better source domains, the classification accuracy will also be good. The experimental results are in Figure~\ref{fig-select}. The results clearly indicated that the proposed USSAR source selection algorithm can choose the right source domains for transfer learning.

\subsection{Evaluation of TNNAR}

In this part, we evaluate the effectiveness of TNNAR for transfer learning in CPAR. In order to test the performance of TNNAR, we conducted two experiments: transfer learning on single source domain, and transfer learning on multiple source domains. We compare TNNAR with the following methods:

\begin{itemize}
	\item PCA: Principal component analysis~\cite{fodor2002survey}.
	\item KPCA: Kernel principal component analysis~\cite{fodor2002survey}.
	\item TCA: Transfer component analysis~\cite{pan2011domain}.
	\item GFK: Geodesic flow kernel~\cite{gong2012geodesic}.
	\item TKL: Transfer kernel learning~\cite{long2015domain}.
	\item STL: Stratified Transfer Learning~\cite{wang2018stratified}.
\end{itemize}

PCA and KPCA are classic dimensionality reduction methods, while TCA, GFK, TKL, and STL are representative transfer learning approaches. The codes of PCA and KPCA are provided in Matlab. The codes of TCA, GFK, and TKL can be obtained online \footnote{\url{https://tinyurl.com/y79j6twy}}. The constructed datasets and STL code are available online~\footnote{\url{https://tinyurl.com/y7en6owt}}.

\begin{table*}[ht]
	\centering
	\caption{Classification accuracy (\%) of TNNAR and other comparison methods on single source transfer tasks.}
	\label{tb-result-single}
	\vspace{-.1in}
	\resizebox{.95\textwidth}{!}{
		\begin{tabular}{|c|c|c|c|c|c|c|c|c|c|}
			\hline
			Scenario & Dataset & \multicolumn{1}{c|}{Task} & \multicolumn{1}{c|}{PCA} & \multicolumn{1}{c|}{KPCA} & \multicolumn{1}{c|}{TCA} & \multicolumn{1}{c|}{GFK} & \multicolumn{1}{c|}{TKL} & \multicolumn{1}{c|}{STL} & \multicolumn{1}{c|}{TNNAR}\\ \hline \hline
			\multirow{4}{*}{Similar body parts on the same person} & \multirow{2}{*}{DSADS} & RA $\rightarrow$ LA & 59.91 & 62.17 & 66.15 & 71.07 & 54.10 & 71.04 & \textbf{75.89} \\ \cline{3-10} 
			&  & RL $\rightarrow$ LL & 69.46 & 70.92 & 75.06 & 79.71 & 61.63 & 81.60  & \textbf{86.76}\\ \cline{2-10} 
			& \multirow{2}{*}{OPP} & RUA $\rightarrow$ LUA & 76.12 & 65.64 & 76.88 & 74.62 & 66.81 & 83.96  & \textbf{87.43}\\ \cline{3-10} 
			&  & RLA $\rightarrow$ LLA & 62.17 & 66.48 & 60.60 & 74.62 & 66.82 & 83.93 & \textbf{86.29} \\ \hline
			\multirow{4}{*}{Different body parts on the same person} & DSADS & RA $\rightarrow$ T & 38.89 & 30.20 & 39.41 & 44.19 & 32.72 & 45.61 & \textbf{50.22}\\ \cline{2-10} 
			& PAMAP2 & H $\rightarrow$ C & 34.97 & 24.44 & 34.86 & 36.24 & 35.67 & 43.47 & \textbf{46.32}\\ \cline{2-10} 
			& \multirow{2}{*}{OPP} & RLA $\rightarrow$ T & 59.10 & 46.99 & 55.43 & 48.89 & 47.66 & 56.88 & \textbf{59.58}\\ \cline{3-10} 
			&  & RUA $\rightarrow$ T & 67.95 & 54.52 & 67.50 & 66.14 & 60.49 & 75.15 & \textbf{75.75}\\ \hline
			\multirow{3}{*}{Similar body parts on different person} & PAMAP2 $\rightarrow$ OPP & C $\rightarrow$ B & 32.80 & 43.78 & 39.02 & 27.64 & 35.64 & 40.10 & \textbf{45.62}\\ \cline{2-10} 
			& DSADS $\rightarrow$ PAMAP & T $\rightarrow$ C & 23.19 & 17.95 & 23.66 & 19.39 & 21.65 & 37.83 & \textbf{39.21}\\ \cline{2-10} 
			& OPP $\rightarrow$ DSADS & B $\rightarrow$ T & 44.30 & 49.35 & 46.91 & 48.07 & 52.79 & 55.45 & \textbf{57.97}\\ \hline \hline
			\multicolumn{3}{|c|}{Average} & 51.71 & 48.40 & 53.23 & 53.69 & 48.73 & 61.37 & \textbf{64.64}\\ \hline
		\end{tabular}
	}
\end{table*}

\begin{table}[htbp]
	\centering
	\caption{Accuracy (\%) of multiple source domains}
	\vspace{-.1in}
	\label{tb-result-multiple}
	\begin{tabular}{|c|c|c|c|c|c|c|}
		\hline
		Target & PCA & TCA & GFK & TKL & STL & TNNAR \\ \hline \hline
		RA & 66.78 & 68.43 & 70.87 & 70.21 & 73.22 & \textbf{78.40} \\ \hline
		Torso & 42.87 & 47.21 & 48.09 & 43.32 & 51.22 & \textbf{55.48} \\ \hline
		RL & 71.24 & 73.47 & 81.23 & 74.26 & 83.76 & \textbf{87.41} \\ \hline
		RLA & 65.78 & 67.10 & 76.38 & 70.32 & 84.52 & \textbf{86.75} \\ \hline \hline
		Average & 61.67 & 64.05 & 69.14 & 64.53 & 73.18 & \textbf{77.01} \\ \hline
	\end{tabular}
\vspace{-.2in}
\end{table}

The implementations of all comparison methods are following~\cite{wang2018stratified}. Different from these work which exploited feature extraction according to human knowledge, we take the original signal as the input. For TNNAR network, we set the learning rate to be $0.001$ with a dropout rate of $0.8$ to prevent overfitting. The batch sizes for source and target domains are 64. Note that although we selected $K$ source domains, we basically combine them into one large source domain. Since the sensor signal is a multi-channel reading, we treat each channel as a distinct signal and perform 1D convolution on it. Totally, there are 9 channels (i.e. 3 accelerometers, 3 gyroscopes, and 3 magnetometers). The convolution kernel size is $64 \times 1$ with the depth $32$. Other parameters of the neural network are set accordingly. For the MMD measurement, we take the linear-time MMD as in~\cite{gretton2012kernel}.

Note that in both of the two experiments, all of the comparison methods used the same source and target domains. For the single source domain situation, we follow the settings in~\cite{wang2018stratified} and report the results in Table~\ref{tb-result-single}. For the experiments on multiple source domains, we extend the results in the last section and set $K=3$ to select the source domains by USSAR. The results are in Table~\ref{tb-result-multiple}. Note that in order to obtain the steady performance, we perform 10 random permutations of the data and record the average results.

The results clearly indicated that TNNAR dramatically increases the performance of cross-domain activity recognition. Specifically, TNNAR has an average performance improvement of \textbf{3.42\%} compared to the best method STL. In all levels of similarities, TNNAR outperforms the comparison methods. It indicates that TNNAR is capable of performing transfer learning in all kinds of activity recognition scenarios. The reasons are three folds: 1) Other comparison methods are operated on the extracted features according to human knowledge, which may not be sufficient to capture the resourceful information of the activities. TNNAR is based on the deep neural network to automatically extract features without human knowledge. As previous work has demonstrated the effectiveness of deep neural network on feature extraction~\cite{wang2017deep}, it will help the network to extract more high-level features. 2) The structure of the neural network is beneficial for performing transfer learning, since the hyperparameters can be easily shared across domains. 3) The deep neural network has both the convolution and LSTM cells, which enables it to learn the spatial and time information from the activities. Therefore, the network can understand more information about the activity data.

We can also have more insights by combining Table~\ref{tb-result-single} and Table~\ref{tb-result-multiple}. Firstly, for the same target domain, adding more source domains will clearly increase the performance. This is because there is more knowledge contained in multiple domains than a single domain. Secondly, when \textit{Torso} is the target domain, the results are dramatically increased. When the arms and legs are as the target domains, the results did not improve that far. This is probably because that the most similar part to the arms and legs are their opposites (the other arms and legs). Thus adding more source domains can only increase the results a little. However, since the Torso is highly correlated with all the body parts, the performance of this body part will be dramatically increased by adding other source domains.

\begin{figure}[]
	\centering
	\vspace{-.1in}
	\hspace{-.1in}
	\subfigure[$\lambda$]{
		\centering
		\includegraphics[scale=0.29]{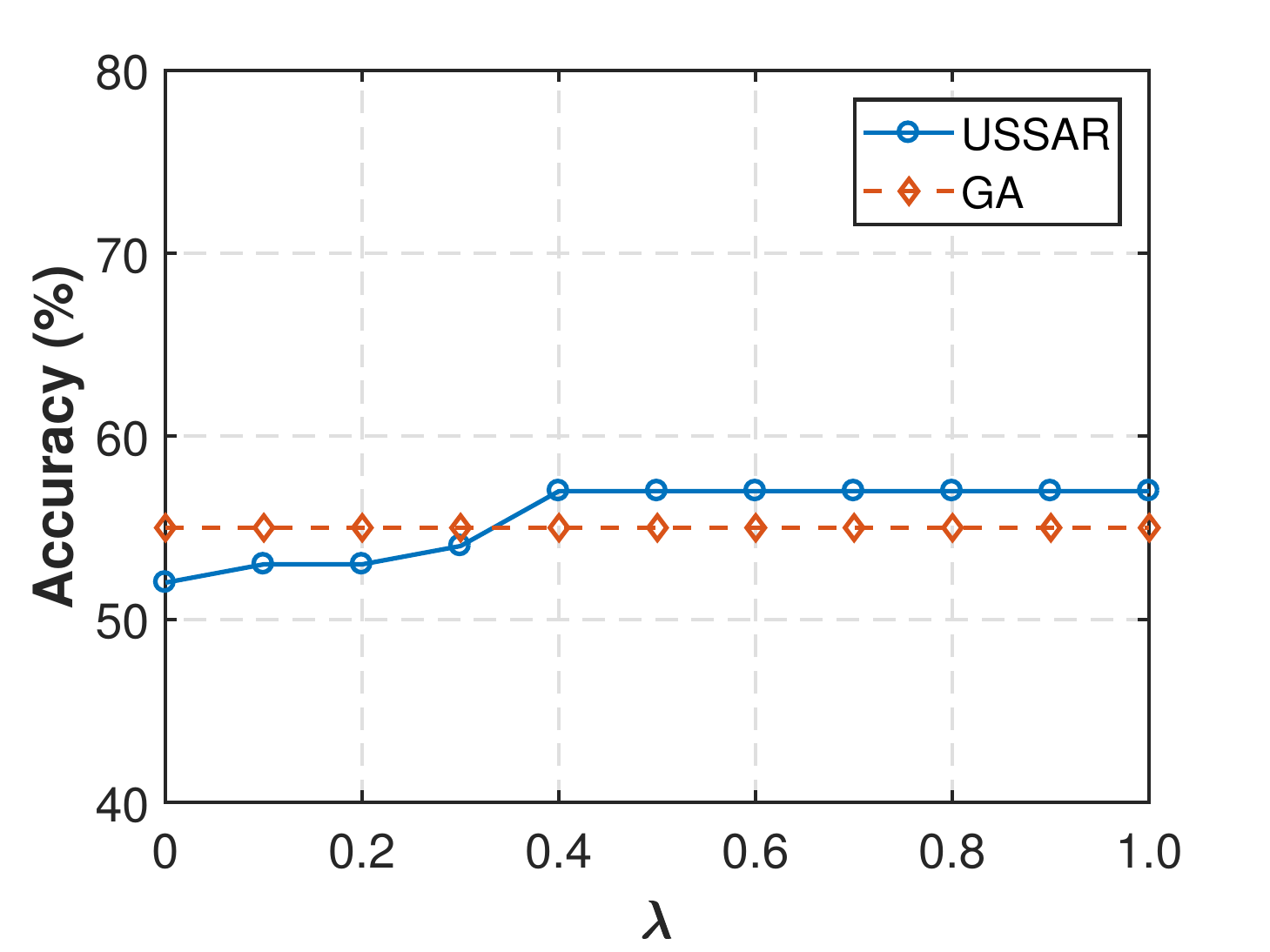}
		\label{fig-sub-lambda}}
	\hspace{-.18in}
	\subfigure[$\mu$]{
		\centering
		\includegraphics[scale=0.29]{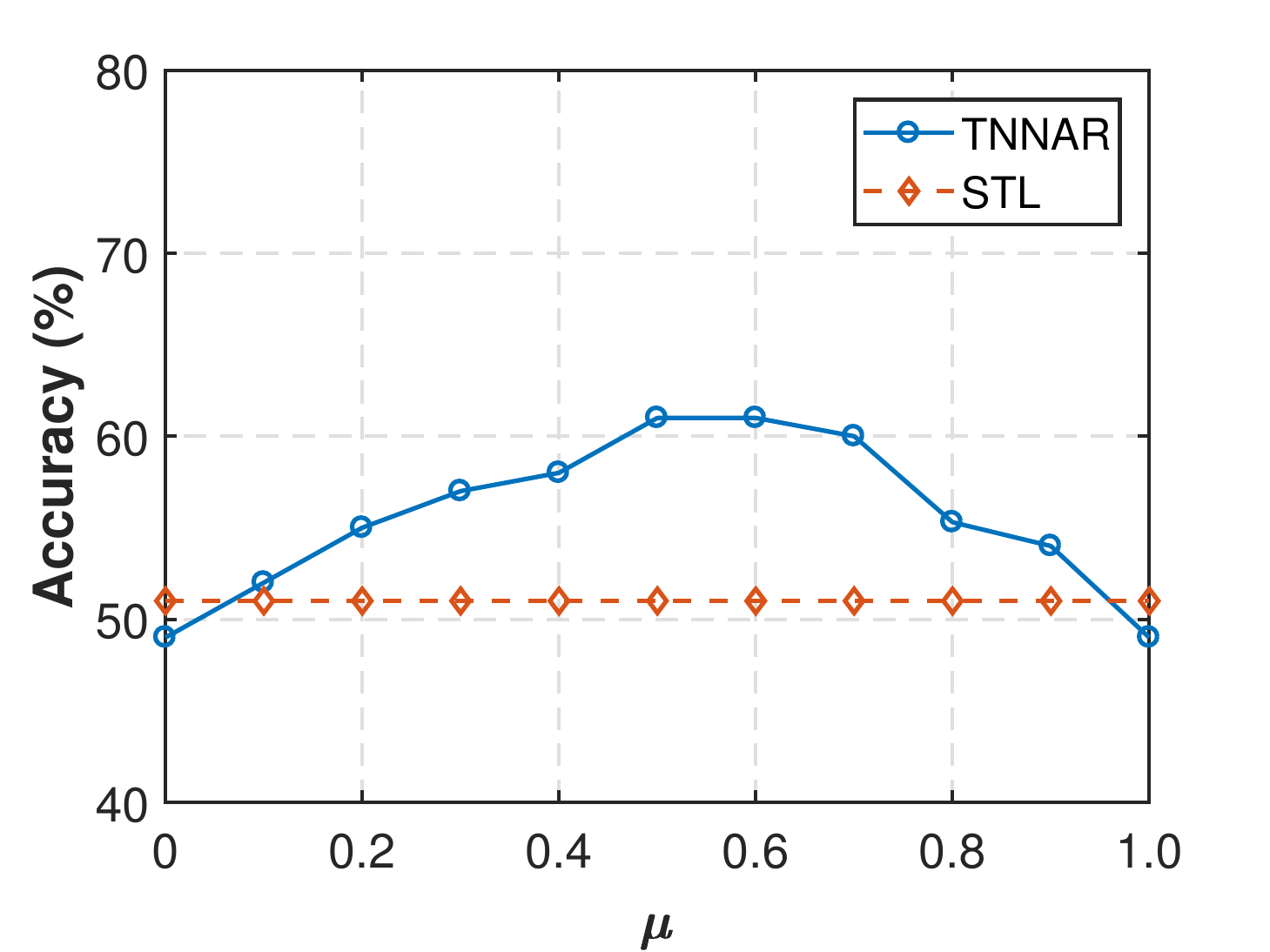}
		\label{fig-sub-mu}}
	\vspace{-.2in}
	\caption{Sensitivity analysis of $\lambda$ and $\mu$.}
	\label{fig-sens}
	\vspace{-.2in}
\end{figure}

\subsection{Sensitivity Analysis}

There are two critical factors in USSAR and TNNAR: the trade-off factors $\lambda$ and $\mu$. A good choice for these two factors will help the algorithm perform better. In this section, we evaluate the sensitivity of them through empirical experiments. We set $\lambda, \mu \in \{0,0.1,\cdots,1.0\}$ for a single task (Torso as the target domain, and $K=3$) and record the classification accuracy in Figure~\ref{fig-sens}. From the results, we can clearly see that the algorithm is robust with a large width of the parameter choice. Therefore, these two parameters do not to be cherry-picked. This indicates that our algorithm can be easily applied to real applications.

\section{Conclusions and Future Work}
\label{sec-con}

If the activity labels of some body parts are missing, it is critical and necessary to exploit the well-labeled information from other body parts to obtain the missing labels. In this paper, we propose the first unsupervised source selection algorithm for activity recognition~(USSAR). USSAR could consider both the semantic and kinetic relation between body parts, thus it is able to accurately select the right domains that are closely related to the target domain. We also propose an end-to-end Transfer Neural Network for activity recognition~(TNNAR) that can learn transferable representations for activities. Experimental results demonstrate that compared to many source selection and transfer learning algorithms, our proposed USSAR can select the right source domains and TNNAR is able to achieve the best classification accuracy.

In future work, we plan to extend the USSAR and TNNAR algorithms for activity recognition with heterogeneous and distant activity domains.

\begin{acks}
	This work is supported by National Key R \& D Program of China (2017YFC0803401) and NSFC (61572471 and 61672313).
\end{acks}

\bibliographystyle{ACM-Reference-Format}
\bibliography{mm18}


\begin{thebibliography}{00}


\ifx \showCODEN    \undefined \def \showCODEN     #1{\unskip}     \fi
\ifx \showDOI      \undefined \def \showDOI       #1{#1}\fi
\ifx \showISBNx    \undefined \def \showISBNx     #1{\unskip}     \fi
\ifx \showISBNxiii \undefined \def \showISBNxiii  #1{\unskip}     \fi
\ifx \showISSN     \undefined \def \showISSN      #1{\unskip}     \fi
\ifx \showLCCN     \undefined \def \showLCCN      #1{\unskip}     \fi
\ifx \shownote     \undefined \def \shownote      #1{#1}          \fi
\ifx \showarticletitle \undefined \def \showarticletitle #1{#1}   \fi
\ifx \showURL      \undefined \def \showURL       {\relax}        \fi
\providecommand\bibfield[2]{#2}
\providecommand\bibinfo[2]{#2}
\providecommand\natexlab[1]{#1}
\providecommand\showeprint[2][]{arXiv:#2}

\bibitem[\protect\citeauthoryear{Barshan and Y{\"u}ksek}{Barshan and
  Y{\"u}ksek}{2014}]%
        {barshan2014recognizing}
\bibfield{author}{\bibinfo{person}{Billur Barshan} {and}
  \bibinfo{person}{Murat~Cihan Y{\"u}ksek}.} \bibinfo{year}{2014}\natexlab{}.
\newblock \showarticletitle{Recognizing daily and sports activities in two open
  source machine learning environments using body-worn sensor units}.
\newblock \bibinfo{journal}{{\it Comput. J.}} \bibinfo{volume}{57},
  \bibinfo{number}{11} (\bibinfo{year}{2014}), \bibinfo{pages}{1649--1667}.
\newblock


\bibitem[\protect\citeauthoryear{Ben-David, Blitzer, Crammer, and
  Pereira}{Ben-David et~al\mbox{.}}{2007}]%
        {ben2007analysis}
\bibfield{author}{\bibinfo{person}{Shai Ben-David}, \bibinfo{person}{John
  Blitzer}, \bibinfo{person}{Koby Crammer}, {and} \bibinfo{person}{Fernando
  Pereira}.} \bibinfo{year}{2007}\natexlab{}.
\newblock \showarticletitle{Analysis of representations for domain adaptation}.
  In \bibinfo{booktitle}{{\em Advances in neural information processing
  systems}}. \bibinfo{pages}{137--144}.
\newblock


\bibitem[\protect\citeauthoryear{Bhatt, Rajkumar, and Roy}{Bhatt
  et~al\mbox{.}}{2016}]%
        {bhatt2016multi}
\bibfield{author}{\bibinfo{person}{Himanshu~S Bhatt}, \bibinfo{person}{Arun
  Rajkumar}, {and} \bibinfo{person}{Shourya Roy}.}
  \bibinfo{year}{2016}\natexlab{}.
\newblock \showarticletitle{Multi-Source Iterative Adaptation for Cross-Domain
  Classification.}. In \bibinfo{booktitle}{{\em IJCAI}}.
  \bibinfo{pages}{3691--3697}.
\newblock


\bibitem[\protect\citeauthoryear{Blitzer, McDonald, and Pereira}{Blitzer
  et~al\mbox{.}}{2006}]%
        {blitzer2006domain}
\bibfield{author}{\bibinfo{person}{John Blitzer}, \bibinfo{person}{Ryan
  McDonald}, {and} \bibinfo{person}{Fernando Pereira}.}
  \bibinfo{year}{2006}\natexlab{}.
\newblock \showarticletitle{Domain adaptation with structural correspondence
  learning}. In \bibinfo{booktitle}{{\em Proceedings of the 2006 conference on
  empirical methods in natural language processing}}. Association for
  Computational Linguistics, \bibinfo{pages}{120--128}.
\newblock


\bibitem[\protect\citeauthoryear{Borgwardt, Gretton, Rasch, Kriegel,
  Sch{\"o}lkopf, and Smola}{Borgwardt et~al\mbox{.}}{2006}]%
        {borgwardt2006integrating}
\bibfield{author}{\bibinfo{person}{Karsten~M Borgwardt},
  \bibinfo{person}{Arthur Gretton}, \bibinfo{person}{Malte~J Rasch},
  \bibinfo{person}{Hans-Peter Kriegel}, \bibinfo{person}{Bernhard
  Sch{\"o}lkopf}, {and} \bibinfo{person}{Alex~J Smola}.}
  \bibinfo{year}{2006}\natexlab{}.
\newblock \showarticletitle{Integrating structured biological data by kernel
  maximum mean discrepancy}.
\newblock \bibinfo{journal}{{\em Bioinformatics\/}} \bibinfo{volume}{22},
  \bibinfo{number}{14} (\bibinfo{year}{2006}), \bibinfo{pages}{e49--e57}.
\newblock


\bibitem[\protect\citeauthoryear{Bulling, Blanke, and Schiele}{Bulling
  et~al\mbox{.}}{2014}]%
        {bulling2014tutorial}
\bibfield{author}{\bibinfo{person}{Andreas Bulling}, \bibinfo{person}{Ulf
  Blanke}, {and} \bibinfo{person}{Bernt Schiele}.}
  \bibinfo{year}{2014}\natexlab{}.
\newblock \showarticletitle{A tutorial on human activity recognition using
  body-worn inertial sensors}.
\newblock \bibinfo{journal}{{\em ACM Computing Surveys (CSUR)\/}}
  \bibinfo{volume}{46}, \bibinfo{number}{3} (\bibinfo{year}{2014}),
  \bibinfo{pages}{33}.
\newblock


\bibitem[\protect\citeauthoryear{Chattopadhyay, Sun, Fan, Davidson,
  Panchanathan, and Ye}{Chattopadhyay et~al\mbox{.}}{2012}]%
        {chattopadhyay2012multisource}
\bibfield{author}{\bibinfo{person}{Rita Chattopadhyay}, \bibinfo{person}{Qian
  Sun}, \bibinfo{person}{Wei Fan}, \bibinfo{person}{Ian Davidson},
  \bibinfo{person}{Sethuraman Panchanathan}, {and} \bibinfo{person}{Jieping
  Ye}.} \bibinfo{year}{2012}\natexlab{}.
\newblock \showarticletitle{Multisource domain adaptation and its application
  to early detection of fatigue}.
\newblock \bibinfo{journal}{{\em ACM Transactions on Knowledge Discovery from
  Data (TKDD)\/}} \bibinfo{volume}{6}, \bibinfo{number}{4}
  (\bibinfo{year}{2012}), \bibinfo{pages}{18}.
\newblock


\bibitem[\protect\citeauthoryear{Chavarriaga, Sagha, Calatroni, Digumarti,
  Tr{\"o}ster, Mill{\'a}n, and Roggen}{Chavarriaga et~al\mbox{.}}{2013}]%
        {chavarriaga2013opportunity}
\bibfield{author}{\bibinfo{person}{Ricardo Chavarriaga}, \bibinfo{person}{Hesam
  Sagha}, \bibinfo{person}{Alberto Calatroni}, \bibinfo{person}{Sundara~Tejaswi
  Digumarti}, \bibinfo{person}{Gerhard Tr{\"o}ster}, \bibinfo{person}{Jos{\'e}
  del~R Mill{\'a}n}, {and} \bibinfo{person}{Daniel Roggen}.}
  \bibinfo{year}{2013}\natexlab{}.
\newblock \showarticletitle{The Opportunity challenge: A benchmark database for
  on-body sensor-based activity recognition}.
\newblock \bibinfo{journal}{{\em Pattern Recognition Letters\/}}
  \bibinfo{volume}{34}, \bibinfo{number}{15} (\bibinfo{year}{2013}),
  \bibinfo{pages}{2033--2042}.
\newblock


\bibitem[\protect\citeauthoryear{Chen, Gu, Jiang, and Wang}{Chen
  et~al\mbox{.}}{2016}]%
        {chen2016ocean}
\bibfield{author}{\bibinfo{person}{Yiqiang Chen}, \bibinfo{person}{Yang Gu},
  \bibinfo{person}{Xinlong Jiang}, {and} \bibinfo{person}{Jindong Wang}.}
  \bibinfo{year}{2016}\natexlab{}.
\newblock \showarticletitle{OCEAN: A New Opportunistic Computing Model for
  Wearable Activity Recognition}. In \bibinfo{booktitle}{{\em Proceedings of
  the 2016 ACM International Joint Conference on Pervasive and Ubiquitous
  Computing: Adjunct}}. \bibinfo{publisher}{ACM}, \bibinfo{pages}{33--36}.
\newblock


\bibitem[\protect\citeauthoryear{Collier, DiBiano, and Mukhopadhyay}{Collier
  et~al\mbox{.}}{2018}]%
        {collier2018cactusnets}
\bibfield{author}{\bibinfo{person}{Edward Collier}, \bibinfo{person}{Robert
  DiBiano}, {and} \bibinfo{person}{Supratik Mukhopadhyay}.}
  \bibinfo{year}{2018}\natexlab{}.
\newblock \showarticletitle{CactusNets: Layer Applicability as a Metric for
  Transfer Learning}.
\newblock \bibinfo{journal}{{\em arXiv preprint arXiv:1804.07846\/}}
  (\bibinfo{year}{2018}).
\newblock


\bibitem[\protect\citeauthoryear{Cook, Feuz, and Krishnan}{Cook
  et~al\mbox{.}}{2013}]%
        {cook2013transfer}
\bibfield{author}{\bibinfo{person}{Diane Cook}, \bibinfo{person}{Kyle~D Feuz},
  {and} \bibinfo{person}{Narayanan~C Krishnan}.}
  \bibinfo{year}{2013}\natexlab{}.
\newblock \showarticletitle{Transfer learning for activity recognition: A
  survey}.
\newblock \bibinfo{journal}{{\em Knowledge and information systems\/}}
  \bibinfo{volume}{36}, \bibinfo{number}{3} (\bibinfo{year}{2013}),
  \bibinfo{pages}{537--556}.
\newblock


\bibitem[\protect\citeauthoryear{Duan, Tsang, and Xu}{Duan
  et~al\mbox{.}}{2012}]%
        {duan2012domain}
\bibfield{author}{\bibinfo{person}{Lixin Duan}, \bibinfo{person}{Ivor~W Tsang},
  {and} \bibinfo{person}{Dong Xu}.} \bibinfo{year}{2012}\natexlab{}.
\newblock \showarticletitle{Domain transfer multiple kernel learning}.
\newblock \bibinfo{journal}{{\em IEEE Transactions on Pattern Analysis and
  Machine Intelligence\/}} \bibinfo{volume}{34}, \bibinfo{number}{3}
  (\bibinfo{year}{2012}), \bibinfo{pages}{465--479}.
\newblock


\bibitem[\protect\citeauthoryear{Fodor}{Fodor}{2002}]%
        {fodor2002survey}
\bibfield{author}{\bibinfo{person}{Imola~K Fodor}.}
  \bibinfo{year}{2002}\natexlab{}.
\newblock \showarticletitle{A survey of dimension reduction techniques}.
\newblock \bibinfo{journal}{{\em Center for Applied Scientific Computing,
  Lawrence Livermore National Laboratory\/}}  \bibinfo{volume}{9}
  (\bibinfo{year}{2002}), \bibinfo{pages}{1--18}.
\newblock


\bibitem[\protect\citeauthoryear{Ganin, Ustinova, Ajakan, Germain, Larochelle,
  Laviolette, Marchand, and Lempitsky}{Ganin et~al\mbox{.}}{2016}]%
        {ganin2016domain}
\bibfield{author}{\bibinfo{person}{Yaroslav Ganin}, \bibinfo{person}{Evgeniya
  Ustinova}, \bibinfo{person}{Hana Ajakan}, \bibinfo{person}{Pascal Germain},
  \bibinfo{person}{Hugo Larochelle}, \bibinfo{person}{Fran{\c{c}}ois
  Laviolette}, \bibinfo{person}{Mario Marchand}, {and} \bibinfo{person}{Victor
  Lempitsky}.} \bibinfo{year}{2016}\natexlab{}.
\newblock \showarticletitle{Domain-adversarial training of neural networks}.
\newblock \bibinfo{journal}{{\em Journal of Machine Learning Research\/}}
  \bibinfo{volume}{17}, \bibinfo{number}{59} (\bibinfo{year}{2016}),
  \bibinfo{pages}{1--35}.
\newblock


\bibitem[\protect\citeauthoryear{Gong, Shi, Sha, and Grauman}{Gong
  et~al\mbox{.}}{2012}]%
        {gong2012geodesic}
\bibfield{author}{\bibinfo{person}{Boqing Gong}, \bibinfo{person}{Yuan Shi},
  \bibinfo{person}{Fei Sha}, {and} \bibinfo{person}{Kristen Grauman}.}
  \bibinfo{year}{2012}\natexlab{}.
\newblock \showarticletitle{Geodesic flow kernel for unsupervised domain
  adaptation}. In \bibinfo{booktitle}{{\em Computer Vision and Pattern
  Recognition (CVPR), 2012 IEEE Conference on}}. IEEE,
  \bibinfo{pages}{2066--2073}.
\newblock


\bibitem[\protect\citeauthoryear{Gretton, Borgwardt, Rasch, Sch{\"o}lkopf, and
  Smola}{Gretton et~al\mbox{.}}{2012}]%
        {gretton2012kernel}
\bibfield{author}{\bibinfo{person}{Arthur Gretton}, \bibinfo{person}{Karsten~M
  Borgwardt}, \bibinfo{person}{Malte~J Rasch}, \bibinfo{person}{Bernhard
  Sch{\"o}lkopf}, {and} \bibinfo{person}{Alexander Smola}.}
  \bibinfo{year}{2012}\natexlab{}.
\newblock \showarticletitle{A kernel two-sample test}.
\newblock \bibinfo{journal}{{\em Journal of Machine Learning Research\/}}
  \bibinfo{volume}{13}, \bibinfo{number}{Mar} (\bibinfo{year}{2012}),
  \bibinfo{pages}{723--773}.
\newblock


\bibitem[\protect\citeauthoryear{Hamm and Lee}{Hamm and Lee}{2008}]%
        {hamm2008grassmann}
\bibfield{author}{\bibinfo{person}{Jihun Hamm} {and} \bibinfo{person}{Daniel~D
  Lee}.} \bibinfo{year}{2008}\natexlab{}.
\newblock \showarticletitle{Grassmann discriminant analysis: a unifying view on
  subspace-based learning}. In \bibinfo{booktitle}{{\em Proceedings of the 25th
  international conference on Machine learning}}. ACM,
  \bibinfo{pages}{376--383}.
\newblock


\bibitem[\protect\citeauthoryear{Hammerla, Fisher, Andras, Rochester, Walker,
  and Pl{\"o}tz}{Hammerla et~al\mbox{.}}{2015}]%
        {hammerla2015pd}
\bibfield{author}{\bibinfo{person}{Nils~Yannick Hammerla},
  \bibinfo{person}{James Fisher}, \bibinfo{person}{Peter Andras},
  \bibinfo{person}{Lynn Rochester}, \bibinfo{person}{Richard Walker}, {and}
  \bibinfo{person}{Thomas Pl{\"o}tz}.} \bibinfo{year}{2015}\natexlab{}.
\newblock \showarticletitle{PD Disease State Assessment in Naturalistic
  Environments Using Deep Learning.}. In \bibinfo{booktitle}{{\em AAAI}}.
  \bibinfo{pages}{1742--1748}.
\newblock


\bibitem[\protect\citeauthoryear{Hossain, Roy, and Khan}{Hossain
  et~al\mbox{.}}{2016}]%
        {hossain2016active}
\bibfield{author}{\bibinfo{person}{HM~Sajjad Hossain},
  \bibinfo{person}{Nirmalya Roy}, {and} \bibinfo{person}{Md~Abdullah Al~Hafiz
  Khan}.} \bibinfo{year}{2016}\natexlab{}.
\newblock \showarticletitle{Active learning enabled activity recognition}. In
  \bibinfo{booktitle}{{\em 2016 IEEE International Conference on Pervasive
  Computing and Communications (PerCom)}}. IEEE, \bibinfo{pages}{1--9}.
\newblock


\bibitem[\protect\citeauthoryear{Hu, Chen, Wang, Wang, Shen, Jiang, and
  Shen}{Hu et~al\mbox{.}}{2016}]%
        {hu2016less}
\bibfield{author}{\bibinfo{person}{Lisha Hu}, \bibinfo{person}{Yiqiang Chen},
  \bibinfo{person}{Shuangquan Wang}, \bibinfo{person}{Jindong Wang},
  \bibinfo{person}{Jianfei Shen}, \bibinfo{person}{Xinlong Jiang}, {and}
  \bibinfo{person}{Zhiqi Shen}.} \bibinfo{year}{2016}\natexlab{}.
\newblock \showarticletitle{Less Annotation on Personalized Activity
  Recognition Using Context Data}. In \bibinfo{booktitle}{{\em Proceedings of
  the 2016 International IEEE Conference on Ubiquitous Intelligence Computing
  (UIC)}}. \bibinfo{pages}{327--332}.
\newblock


\bibitem[\protect\citeauthoryear{Lara and Labrador}{Lara and Labrador}{2013}]%
        {lara2013survey}
\bibfield{author}{\bibinfo{person}{Oscar~D Lara} {and}
  \bibinfo{person}{Miguel~A Labrador}.} \bibinfo{year}{2013}\natexlab{}.
\newblock \showarticletitle{A survey on human activity recognition using
  wearable sensors}.
\newblock \bibinfo{journal}{{\em IEEE Communications Surveys \& Tutorials\/}}
  \bibinfo{volume}{15}, \bibinfo{number}{3} (\bibinfo{year}{2013}),
  \bibinfo{pages}{1192--1209}.
\newblock


\bibitem[\protect\citeauthoryear{Lasecki, Song, Kautz, and Bigham}{Lasecki
  et~al\mbox{.}}{2013}]%
        {lasecki2013real}
\bibfield{author}{\bibinfo{person}{Walter~S Lasecki},
  \bibinfo{person}{Young~Chol Song}, \bibinfo{person}{Henry Kautz}, {and}
  \bibinfo{person}{Jeffrey~P Bigham}.} \bibinfo{year}{2013}\natexlab{}.
\newblock \showarticletitle{Real-time crowd labeling for deployable activity
  recognition}. In \bibinfo{booktitle}{{\em Proceedings of the 2013 conference
  on Computer supported cooperative work}}. ACM, \bibinfo{pages}{1203--1212}.
\newblock


\bibitem[\protect\citeauthoryear{LeCun, Bengio, and Hinton}{LeCun
  et~al\mbox{.}}{2015}]%
        {lecun2015deep}
\bibfield{author}{\bibinfo{person}{Yann LeCun}, \bibinfo{person}{Yoshua
  Bengio}, {and} \bibinfo{person}{Geoffrey Hinton}.}
  \bibinfo{year}{2015}\natexlab{}.
\newblock \showarticletitle{Deep learning}.
\newblock \bibinfo{journal}{{\em Nature\/}} \bibinfo{volume}{521},
  \bibinfo{number}{7553} (\bibinfo{year}{2015}), \bibinfo{pages}{436--444}.
\newblock


\bibitem[\protect\citeauthoryear{Long, Wang, Ding, Sun, and Yu}{Long
  et~al\mbox{.}}{2013}]%
        {long2013transfer}
\bibfield{author}{\bibinfo{person}{Mingsheng Long}, \bibinfo{person}{Jianmin
  Wang}, \bibinfo{person}{Guiguang Ding}, \bibinfo{person}{Jiaguang Sun}, {and}
  \bibinfo{person}{Philip~S Yu}.} \bibinfo{year}{2013}\natexlab{}.
\newblock \showarticletitle{Transfer feature learning with joint distribution
  adaptation}. In \bibinfo{booktitle}{{\em Proceedings of the IEEE
  International Conference on Computer Vision}}. \bibinfo{pages}{2200--2207}.
\newblock


\bibitem[\protect\citeauthoryear{Long, Wang, Sun, and Philip}{Long
  et~al\mbox{.}}{2015}]%
        {long2015domain}
\bibfield{author}{\bibinfo{person}{Mingsheng Long}, \bibinfo{person}{Jianmin
  Wang}, \bibinfo{person}{Jiaguang Sun}, {and} \bibinfo{person}{S~Yu Philip}.}
  \bibinfo{year}{2015}\natexlab{}.
\newblock \showarticletitle{Domain invariant transfer kernel learning}.
\newblock \bibinfo{journal}{{\em IEEE Transactions on Knowledge and Data
  Engineering\/}} \bibinfo{volume}{27}, \bibinfo{number}{6}
  (\bibinfo{year}{2015}), \bibinfo{pages}{1519--1532}.
\newblock


\bibitem[\protect\citeauthoryear{Long, Zhu, Wang, and Jordan}{Long
  et~al\mbox{.}}{2017}]%
        {long2017deep}
\bibfield{author}{\bibinfo{person}{Mingsheng Long}, \bibinfo{person}{Han Zhu},
  \bibinfo{person}{Jianmin Wang}, {and} \bibinfo{person}{Michael~I Jordan}.}
  \bibinfo{year}{2017}\natexlab{}.
\newblock \showarticletitle{Deep Transfer Learning with Joint Adaptation
  Networks}. In \bibinfo{booktitle}{{\em International Conference on Machine
  Learning}}. \bibinfo{pages}{2208--2217}.
\newblock


\bibitem[\protect\citeauthoryear{Lu, Zhu, Pan, Xiang, Wang, and Yang}{Lu
  et~al\mbox{.}}{2014}]%
        {lu2014source}
\bibfield{author}{\bibinfo{person}{Zhongqi Lu}, \bibinfo{person}{Yin Zhu},
  \bibinfo{person}{Sinno~Jialin Pan}, \bibinfo{person}{Evan~Wei Xiang},
  \bibinfo{person}{Yujing Wang}, {and} \bibinfo{person}{Qiang Yang}.}
  \bibinfo{year}{2014}\natexlab{}.
\newblock \showarticletitle{Source Free Transfer Learning for Text
  Classification.}. In \bibinfo{booktitle}{{\em AAAI}}.
  \bibinfo{pages}{122--128}.
\newblock


\bibitem[\protect\citeauthoryear{Morales and Roggen}{Morales and
  Roggen}{2016}]%
        {morales2016deep}
\bibfield{author}{\bibinfo{person}{Francisco Javier~Ord{\'o}{\~n}ez Morales}
  {and} \bibinfo{person}{Daniel Roggen}.} \bibinfo{year}{2016}\natexlab{}.
\newblock \showarticletitle{Deep convolutional feature transfer across mobile
  activity recognition domains, sensor modalities and locations}. In
  \bibinfo{booktitle}{{\em Proceedings of the 2016 ACM International Symposium
  on Wearable Computers}}. ACM, \bibinfo{pages}{92--99}.
\newblock


\bibitem[\protect\citeauthoryear{Nguyen, Zeng, Tague, and Zhang}{Nguyen
  et~al\mbox{.}}{2015}]%
        {nguyen2015did}
\bibfield{author}{\bibinfo{person}{Le~T Nguyen}, \bibinfo{person}{Ming Zeng},
  \bibinfo{person}{Patrick Tague}, {and} \bibinfo{person}{Joy Zhang}.}
  \bibinfo{year}{2015}\natexlab{}.
\newblock \showarticletitle{I did not smoke 100 cigarettes today!: avoiding
  false positives in real-world activity recognition}. In
  \bibinfo{booktitle}{{\em Proceedings of the 2015 ACM International Joint
  Conference on Pervasive and Ubiquitous Computing}}. ACM,
  \bibinfo{pages}{1053--1063}.
\newblock


\bibitem[\protect\citeauthoryear{Pan, Kwok, and Yang}{Pan
  et~al\mbox{.}}{2008}]%
        {pan2008transfer}
\bibfield{author}{\bibinfo{person}{Sinno~Jialin Pan}, \bibinfo{person}{James~T
  Kwok}, {and} \bibinfo{person}{Qiang Yang}.} \bibinfo{year}{2008}\natexlab{}.
\newblock \showarticletitle{Transfer Learning via Dimensionality Reduction}. In
  \bibinfo{booktitle}{{\em Proceedings of the 23rd AAAI conference on
  Artificial intelligence}}, Vol.~\bibinfo{volume}{8}.
  \bibinfo{pages}{677--682}.
\newblock


\bibitem[\protect\citeauthoryear{Pan, Tsang, Kwok, and Yang}{Pan
  et~al\mbox{.}}{2011}]%
        {pan2011domain}
\bibfield{author}{\bibinfo{person}{Sinno~Jialin Pan}, \bibinfo{person}{Ivor~W
  Tsang}, \bibinfo{person}{James~T Kwok}, {and} \bibinfo{person}{Qiang Yang}.}
  \bibinfo{year}{2011}\natexlab{}.
\newblock \showarticletitle{Domain adaptation via transfer component analysis}.
\newblock \bibinfo{journal}{{\em IEEE Transactions on Neural Networks\/}}
  \bibinfo{volume}{22}, \bibinfo{number}{2} (\bibinfo{year}{2011}),
  \bibinfo{pages}{199--210}.
\newblock


\bibitem[\protect\citeauthoryear{Pan and Yang}{Pan and Yang}{2010}]%
        {pan2010survey}
\bibfield{author}{\bibinfo{person}{Sinno~Jialin Pan} {and}
  \bibinfo{person}{Qiang Yang}.} \bibinfo{year}{2010}\natexlab{}.
\newblock \showarticletitle{A survey on transfer learning}.
\newblock \bibinfo{journal}{{\em Knowledge and Data Engineering, IEEE
  Transactions on\/}} \bibinfo{volume}{22}, \bibinfo{number}{10}
  (\bibinfo{year}{2010}), \bibinfo{pages}{1345--1359}.
\newblock


\bibitem[\protect\citeauthoryear{Pl{\"o}tz, Hammerla, and Olivier}{Pl{\"o}tz
  et~al\mbox{.}}{2011}]%
        {plotz2011feature}
\bibfield{author}{\bibinfo{person}{Thomas Pl{\"o}tz}, \bibinfo{person}{Nils~Y
  Hammerla}, {and} \bibinfo{person}{Patrick Olivier}.}
  \bibinfo{year}{2011}\natexlab{}.
\newblock \showarticletitle{Feature learning for activity recognition in
  ubiquitous computing}. In \bibinfo{booktitle}{{\em IJCAI
  Proceedings-International Joint Conference on Artificial Intelligence}},
  Vol.~\bibinfo{volume}{22}. \bibinfo{pages}{1729}.
\newblock


\bibitem[\protect\citeauthoryear{Reiss and Stricker}{Reiss and
  Stricker}{2012}]%
        {reiss2012introducing}
\bibfield{author}{\bibinfo{person}{Attila Reiss} {and} \bibinfo{person}{Didier
  Stricker}.} \bibinfo{year}{2012}\natexlab{}.
\newblock \showarticletitle{Introducing a new benchmarked dataset for activity
  monitoring}. In \bibinfo{booktitle}{{\em Wearable Computers (ISWC), 2012 16th
  International Symposium on}}. IEEE, \bibinfo{pages}{108--109}.
\newblock


\bibitem[\protect\citeauthoryear{Sung, Yang, Zhang, Xiang, Torr, and
  Hospedales}{Sung et~al\mbox{.}}{2017}]%
        {sung2017learning}
\bibfield{author}{\bibinfo{person}{Flood Sung}, \bibinfo{person}{Yongxin Yang},
  \bibinfo{person}{Li Zhang}, \bibinfo{person}{Tao Xiang},
  \bibinfo{person}{Philip~HS Torr}, {and} \bibinfo{person}{Timothy~M
  Hospedales}.} \bibinfo{year}{2017}\natexlab{}.
\newblock \showarticletitle{Learning to Compare: Relation Network for Few-Shot
  Learning}.
\newblock \bibinfo{journal}{{\em arXiv preprint arXiv:1711.06025\/}}
  (\bibinfo{year}{2017}).
\newblock


\bibitem[\protect\citeauthoryear{Tan, Zhang, Pan, and Yang}{Tan
  et~al\mbox{.}}{2017}]%
        {tan2017distant}
\bibfield{author}{\bibinfo{person}{Ben Tan}, \bibinfo{person}{Yu Zhang},
  \bibinfo{person}{Sinno~Jialin Pan}, {and} \bibinfo{person}{Qiang Yang}.}
  \bibinfo{year}{2017}\natexlab{}.
\newblock \showarticletitle{Distant Domain Transfer Learning}. In
  \bibinfo{booktitle}{{\em Thirty-First AAAI Conference on Artificial
  Intelligence}}.
\newblock


\bibitem[\protect\citeauthoryear{Tzeng, Hoffman, Zhang, Saenko, and
  Darrell}{Tzeng et~al\mbox{.}}{2014}]%
        {tzeng2014deep}
\bibfield{author}{\bibinfo{person}{Eric Tzeng}, \bibinfo{person}{Judy Hoffman},
  \bibinfo{person}{Ning Zhang}, \bibinfo{person}{Kate Saenko}, {and}
  \bibinfo{person}{Trevor Darrell}.} \bibinfo{year}{2014}\natexlab{}.
\newblock \showarticletitle{Deep domain confusion: Maximizing for domain
  invariance}.
\newblock \bibinfo{journal}{{\em arXiv preprint arXiv:1412.3474\/}}
  (\bibinfo{year}{2014}).
\newblock


\bibitem[\protect\citeauthoryear{Wang, Chen, Hao, Feng, and Shen}{Wang
  et~al\mbox{.}}{2017}]%
        {wang2017balanced}
\bibfield{author}{\bibinfo{person}{Jindong Wang}, \bibinfo{person}{Yiqiang
  Chen}, \bibinfo{person}{Shuji Hao}, \bibinfo{person}{Wenjie Feng}, {and}
  \bibinfo{person}{Zhiqi Shen}.} \bibinfo{year}{2017}\natexlab{}.
\newblock \showarticletitle{Balanced Distribution Adaptation for Transfer
  Learning}. In \bibinfo{booktitle}{{\em The IEEE International conference on
  data mining (ICDM)}}. \bibinfo{pages}{1129--1134}.
\newblock


\bibitem[\protect\citeauthoryear{Wang, Chen, Hao, Peng, and Hu}{Wang
  et~al\mbox{.}}{2018a}]%
        {wang2017deep}
\bibfield{author}{\bibinfo{person}{Jindong Wang}, \bibinfo{person}{Yiqiang
  Chen}, \bibinfo{person}{Shuji Hao}, \bibinfo{person}{Xiaohui Peng}, {and}
  \bibinfo{person}{Lisha Hu}.} \bibinfo{year}{2018}\natexlab{a}.
\newblock \showarticletitle{Deep Learning for Sensor-based Activity
  Recognition: A Survey}.
\newblock \bibinfo{journal}{{\em Pattern Recognition Letters\/}}
  (\bibinfo{year}{2018}).
\newblock


\bibitem[\protect\citeauthoryear{Wang, Chen, Hu, Peng, and Yu}{Wang
  et~al\mbox{.}}{2018b}]%
        {wang2018stratified}
\bibfield{author}{\bibinfo{person}{Jindong Wang}, \bibinfo{person}{Yiqiang
  Chen}, \bibinfo{person}{Lisha Hu}, \bibinfo{person}{Xiaohui Peng}, {and}
  \bibinfo{person}{Philip~S Yu}.} \bibinfo{year}{2018}\natexlab{b}.
\newblock \showarticletitle{Stratified Transfer Learning for Cross-domain
  Activity Recognition}. In \bibinfo{booktitle}{{\em IEEE international
  conference on pervasive computing and communications (PerCom)}}.
\newblock


\bibitem[\protect\citeauthoryear{Wen, Indulska, and Zhong}{Wen
  et~al\mbox{.}}{2016}]%
        {wen2016adaptive}
\bibfield{author}{\bibinfo{person}{Jiahui Wen}, \bibinfo{person}{Jadwiga
  Indulska}, {and} \bibinfo{person}{Mingyang Zhong}.}
  \bibinfo{year}{2016}\natexlab{}.
\newblock \showarticletitle{Adaptive activity learning with dynamically
  available context}. In \bibinfo{booktitle}{{\em 2016 IEEE International
  Conference on Pervasive Computing and Communications (PerCom)}}. IEEE,
  \bibinfo{pages}{1--11}.
\newblock


\bibitem[\protect\citeauthoryear{Xiang, Pan, Pan, Su, and Yang}{Xiang
  et~al\mbox{.}}{2011}]%
        {xiang2011source}
\bibfield{author}{\bibinfo{person}{Evan~Wei Xiang},
  \bibinfo{person}{Sinno~Jialin Pan}, \bibinfo{person}{Weike Pan},
  \bibinfo{person}{Jian Su}, {and} \bibinfo{person}{Qiang Yang}.}
  \bibinfo{year}{2011}\natexlab{}.
\newblock \showarticletitle{Source-selection-free transfer learning}. In
  \bibinfo{booktitle}{{\em IJCAI proceedings-international joint conference on
  artificial intelligence}}, Vol.~\bibinfo{volume}{22}. \bibinfo{pages}{2355}.
\newblock


\bibitem[\protect\citeauthoryear{Xu, Yang, Zhou, Shangguan, Yi, and Liu}{Xu
  et~al\mbox{.}}{2016}]%
        {xu2016indoor}
\bibfield{author}{\bibinfo{person}{Han Xu}, \bibinfo{person}{Zheng Yang},
  \bibinfo{person}{Zimu Zhou}, \bibinfo{person}{Longfei Shangguan},
  \bibinfo{person}{Ke Yi}, {and} \bibinfo{person}{Yunhao Liu}.}
  \bibinfo{year}{2016}\natexlab{}.
\newblock \showarticletitle{Indoor localization via multi-modal sensing on
  smartphones}. In \bibinfo{booktitle}{{\em UbiComp}}. ACM,
  \bibinfo{pages}{208--219}.
\newblock


\bibitem[\protect\citeauthoryear{Yao and Doretto}{Yao and Doretto}{2010}]%
        {yao2010boosting}
\bibfield{author}{\bibinfo{person}{Yi Yao} {and} \bibinfo{person}{Gianfranco
  Doretto}.} \bibinfo{year}{2010}\natexlab{}.
\newblock \showarticletitle{Boosting for transfer learning with multiple
  sources}. In \bibinfo{booktitle}{{\em Computer vision and pattern recognition
  (CVPR), 2010 IEEE conference on}}. IEEE, \bibinfo{pages}{1855--1862}.
\newblock


\bibitem[\protect\citeauthoryear{Yosinski, Clune, Bengio, and Lipson}{Yosinski
  et~al\mbox{.}}{2014}]%
        {yosinski2014transferable}
\bibfield{author}{\bibinfo{person}{Jason Yosinski}, \bibinfo{person}{Jeff
  Clune}, \bibinfo{person}{Yoshua Bengio}, {and} \bibinfo{person}{Hod Lipson}.}
  \bibinfo{year}{2014}\natexlab{}.
\newblock \showarticletitle{How transferable are features in deep neural
  networks?}. In \bibinfo{booktitle}{{\em Advances in neural information
  processing systems}}. \bibinfo{pages}{3320--3328}.
\newblock


\bibitem[\protect\citeauthoryear{Zhao, Yue, Katabi, and Jaakkola}{Zhao
  et~al\mbox{.}}{2017}]%
        {zhao2017sleep}
\bibfield{author}{\bibinfo{person}{Ming Zhao}, \bibinfo{person}{Shichao Yue},
  \bibinfo{person}{Dina Katabi}, {and} \bibinfo{person}{Tommi Jaakkola}.}
  \bibinfo{year}{2017}\natexlab{}.
\newblock \showarticletitle{Learning sleep stages from radio signals: A deep
  adversarial architecture}. In \bibinfo{booktitle}{{\em ICML}}.
\newblock


\bibitem[\protect\citeauthoryear{Zhao, Chen, Liu, Shen, and Liu}{Zhao
  et~al\mbox{.}}{2011}]%
        {zhao2011cross}
\bibfield{author}{\bibinfo{person}{Zhongtang Zhao}, \bibinfo{person}{Yiqiang
  Chen}, \bibinfo{person}{Junfa Liu}, \bibinfo{person}{Zhiqi Shen}, {and}
  \bibinfo{person}{Mingjie Liu}.} \bibinfo{year}{2011}\natexlab{}.
\newblock \showarticletitle{Cross-people mobile-phone based activity
  recognition}. In \bibinfo{booktitle}{{\em Proceedings of the Twenty-Second
  international joint conference on Artificial Intelligence (IJCAI)}},
  Vol.~\bibinfo{volume}{11}. Citeseer, \bibinfo{pages}{2545--2550}.
\newblock


\bibitem[\protect\citeauthoryear{Zheng and Yang}{Zheng and Yang}{2011}]%
        {zheng2011user}
\bibfield{author}{\bibinfo{person}{Vincent~W Zheng} {and}
  \bibinfo{person}{Qiang Yang}.} \bibinfo{year}{2011}\natexlab{}.
\newblock \showarticletitle{User-dependent aspect model for collaborative
  activity recognition}. In \bibinfo{booktitle}{{\em Proceedings of the
  Twenty-Second international joint conference on Artificial Intelligence
  (IJCAI)}}, Vol.~\bibinfo{volume}{22}. \bibinfo{pages}{2085--2090}.
\newblock


\end{thebibliography}

\end{document}